\documentclass[
aps,%
12pt,%
final,%
notitlepage,%
oneside,%
onecolumn,%
nobibnotes,%
nofootinbib,%
superscriptaddress,%
showpacs,%
]%
{revtex4}
\usepackage{epsfig}
\usepackage{graphics}
\def\Ref#1{(\ref{#1})}
\newcommand{\be}{\begin{equation}}
\newcommand{\ee}{\end{equation}}
\newcommand{\bn}{\begin{eqnarray}}
\newcommand{\en}{\end{eqnarray}}
\usepackage[english]{babel}
\begin{document}

\title{Diffusion Time-Scale Invariance, Markovization Processes and Memory Effects in Lennard-Jones Liquids}

\author{\firstname{Renat~M.}~\surname{Yulmetyev}}
\email{rmy@dtp.ksu.ras.ru} \affiliation{Department of Physics,
Kazan State Pedagogical University, 420021 Kazan, Russia}

\author{\firstname{Anatolii~V.}~\surname{Mokshin}}
\email{mav@dtp.ksu.ras.ru} \affiliation{Department of Physics,
Kazan State Pedagogical University, 420021 Kazan, Russia}

\author{\firstname{Peter}~\surname{H\"anggi}}
\affiliation{Department of Physics, University of Augsburg,
D-86135 Augsburg, Germany}

\today
\begin{abstract}
We report the results of calculation of diffusion coefficients for
Lennard-Jones liquids, based on the idea of time-scale invariance
of relaxation processes in liquids. The results were compared with
the molecular dynamics data for  Lennard-Jones system and a good
agreement of our theory with these data over a wide range of
densities and temperatures was obtained. By calculations of the
non-Markovity parameter we have estimated numerically statistical
memory effects of diffusion in  detail.
\end{abstract}

 \maketitle


The motion of atoms and molecules in liquids is a typical example
of a many-body problem. It is well known, however, that in
contrast to gases and solids the physical mechanisms of transport
processes in liquid matter is no well established \cite{Balucani}.
In particular, for diffusive systems the role of memory effects
induced by the disorder of the background medium is not
investigated enough \cite{Hanggi,Morgado}. There are different
theoretical approaches and approximations for calculating
transport coefficients \cite{Balucani,Boon}, in some of them they
resort to time correlation functions  (see, for example,
\cite{Balucani,Boon,Hansen,Tankeshwar,Morgado}). The general
approach of time  correlation functions leads to calculation of
exact expressions of transport coefficients for hydrodynamic
equations. The similar expressions are known as the Green-Kubo
relations.

In the present work we suggest a new  approach to calculating
transport properties based on the realization of idea of
time-scale invariance of relaxation processes  in liquids
\cite{Invariance} by well-known Zwanzig-Mori's memory function
formalism \cite{Zwanzig,Fuchs}. Recently this idea has made it
possible to explain the experimental data on slow neutron
scattering  in liquid cesium  and sodium \cite{Invariance}. Here
we have tested this approach on Lennard-Jones (LJ) liquids. As
known, among a large set of dense fluids these systems are one of
the most popular (base) for different investigations. So,
diffusion coefficients of LJ liquids were calculated for a wide
range of densities and temperatures, and then quantitative
analysis of the memory effects in diffusion phenomena has
executed. The obtained results are compared with the predictions
of other theories \cite{Tankeshwar} and the molecular dynamics
data \cite{Heyes}.


As known, the diffusion coefficient can be expressed in terms of
the velocity autocorrelation function (VACF)

\be a(t)=\frac{\langle v_{\alpha}(0) v_{\alpha}(t)
\rangle}{\langle v_{\alpha}(0)^{2} \rangle}
\ee

\noindent by the relation

\be D=\frac{k_{B}T}{m}\int_{0}^{\infty}a(t)dt. \label{diff} \ee

\noindent The latter is one of the Green-Kubo equations, relating
the integrals of the correlation function to the transport
coefficients \cite{Balucani}. Here $k_{B}$, $T$ and $m$ are the
Boltzmann constant, temperature and atomic mass, respectively.

On the other side, the generalized Langevin equation (GLE) for
VACF can be written in the following way:

\be \frac{da(t)}{dt}=-\omega^{(2)} \int_{0}^{t}
M_{1}(\tau)a(t-\tau)d\tau, \ee

\noindent where $M_{1}(t)$ is the first-order normalized memory
function $[M_{1}(t=0)=1]$ and $\omega^{(2)}$ is the second
frequency moment of  VACF. This equation is obtained by the use of
Zwanzig-Mori's projection operators formalism \cite{Zwanzig}.
However, the projection operators technique allows to derive the
whole chain of interconnected equations similar to the above
mentioned equation. These equations  contain the memory functions
$M_{i}(t)$ and the even frequency moments $\omega^{(2i)}$ of
higher orders, i.e. $i=1,2,3,...$. This chain can be represented
in terms of Laplace transforms as a continued fraction:

\bn \tilde a(s)&=&\int^{\infty}_{0}dt
e^{-st}a(t)=[s+\omega^{(2)}\tilde{M}_{1}(s)]^{-1}=1/\left \lbrace
s+\omega^{(2)}/\left \lbrack
s+(\omega^{(4)}/\omega^{(2)}-\omega^{(2)})\tilde{M}_{2}(s)\right
\rbrack
\right \rbrace \nonumber\\
&=&1/\left \lbrace s+\omega^{(2)}/\left \lbrack
s+(\omega^{(4)}/\omega^{(2)}-\omega^{(2)})/(s+...)\right \rbrack
\right \rbrace, \label{con_frac} \en

\noindent and Eq. \Ref{diff} yields

\be
D=\frac{k_{B}T}{m}\tilde {a}(s=0). \label{general}
\ee

So, the problem of calculation of transport properties can be
reduced to  calculation of the certain memory functions
$M_{i}(t)$. Although there are microscopic expressions and some
formal prescriptions for  memory functions calculations,  it is
prohibitively difficult for models like  Lennard-Jones (LJ)
liquid. A rich variety of model functions was proposed for this
purpose, most of which have little physical justification. For
example, hyperbolic secant memory was used to calculate velocity,
transverse stress , energy current density correlation functions
and the corresponding transport coefficients for LJ liquid
\cite{Boon}. The validity and justification of this approach was
closely examined in work \cite{Lee}.

There is another more effective and powerful way to calculate
transport coefficients of the diffusion constant type. This method
was first suggested in work \cite{Yulmetyev76} to calculate
self-diffusion in liquid argon, and used later in Ref.
\cite{Shurygin} to investigate viscosity effects in liquid argon.
Here we use this approach to analyze  diffusion time-scales and to
calculate the diffusion coefficient in LJ liquid over a wide
density and temperature ranges.

As known, the memory functions $M_{i}(t)$ have characteristic time
scales, which can generally be defined at fixed $i$  by
$\tau_{i}=\tilde{M}_{i}(s=0)=\int^{\infty}_{0}dt M_{i}(t)$, where
$\tau_{i}$ is the relaxation time of a certain  correlation
function $M_{i}(t)$.  These time scales characterize the
corresponding relaxation processes and can have different
numerical values. Nonetheless, on a certain relaxation level, (for
example, on the $i$th level) the scale invariance of the nearest
interconnected relaxation processes can exist.  We shall consider
below the following approximation $\tau_{i+1}=\tau_{i}$.
Physically it means the occurrence of time-scale invariance on the
$i$th relaxation level.

In  case of $i=0$ we have an approximate equality of VACF
relaxation time $\tau_{0}$ and the relaxation time of the
first-order memory function $\tau_{1}$. Then, from Eq.
\Ref{general} and the second equality of Eq. \Ref{con_frac} at
$\tau_{1}=\tau_{0}$ one can obtain

\be D=\frac{k_{B}T}{m} \left \lbrack \frac{1}{\omega^{(2)}}\right
\rbrack^{1/2},\ \omega^{(2)}=\frac{4\pi n}{3}\int_{0}^{\infty}dr
g(r) r^{2} \left \lbrack \frac{3}{r}\frac{\partial U(r)}{\partial
r}+r\frac{\partial}{\partial r}\left(\frac{\partial
U(r)}{r\partial r}\right)\right \rbrack, \label{first} \ee where
$n$ is  density, $g(r)$ is radial distribution function and $U(r)$
is interparticle potential.

\begin{figure}
\centerline{\epsfig{figure=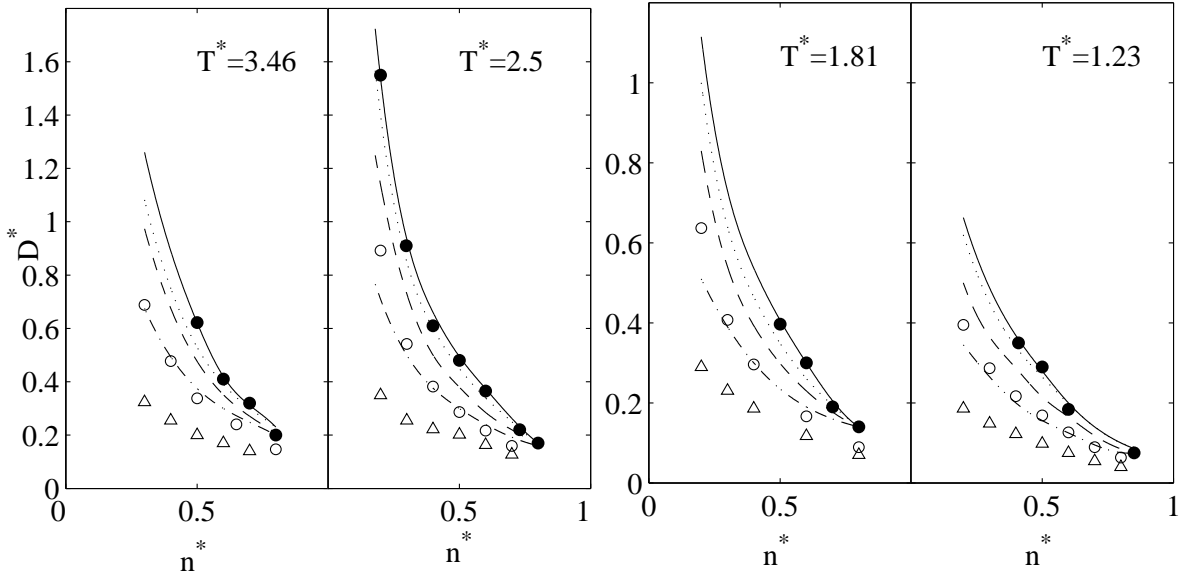,height=8cm,angle=0}}
\caption{The reduced density $n^{*}=n\sigma^{3}$ dependence of
diffusion coefficient at the reduced temperature
$T^{*}=k_{B}T/\varepsilon=3.46$, $2.5$, $1.81$ and $1.23$. The
triangles ($\triangle$ $\triangle$ $\triangle$) show results of
Eq. \Ref{first}, the solid line presents our results from Eq.
\Ref{second}, the circles ($\circ$ $\circ$ $\circ$) show results
of Eq. \Ref{third}, the dotted line ($\cdots$) presents the
results of approximation with hyperbolic secant memory of Ref.
\cite{Tankeshwar}, the broken ($---$) and the chain line ($- \cdot
-$) correspond to results of two models for diffusion coefficient
obtained on the basis of Joslin and Gray study in terms of the
first two and three Mori's coefficients, correspondingly
\cite{Tankeshwar,Joslin}. The full circles ($\bullet$) represent
molecular dynamics data of Ref. \cite{Heyes} and correspond to
diffusion values in points well within the fluid region of the LJ
phase diagram \cite{mds}. Presented data bring out clearly  that
results of our theory are in excellent agreement with molecular
dynamics data without any deviations.}
\end{figure}

At $i=1$ we obtain the equality of the relaxation times of first
and  second order memory functions \cite{expon}. In this case,
$\tau_{2}=\tau_{1}$, and from Eqs. \Ref{con_frac} and
\Ref{general} we have

\bn D&=&\frac{k_{B}T}{m}\frac{1}{\omega^{(2)}} \left \lbrack
\frac{\omega^{(4)}}{\omega^{(2)}}-\omega^{(2)} \right
\rbrack^{1/2},\nonumber\\
\omega^{(4)}&=&\frac{8\pi n}{3m}\int_{0}^{\infty}dr g(r)\left [
3\left (\frac{dU(r)}{dr}\right )^{2}+\left
(r\frac{\partial}{\partial r}\left(\frac{\partial U(r)}{r\partial
r}\right)\right )^{2}+  \frac{\partial U(r)}{\partial r}\cdot
\frac{\partial}{\partial r}\left(\frac{\partial U(r)}{r\partial
r}\right)
\right ]\nonumber\\
& &+\frac{8\pi^{2}n^{2}}{3m}\int_{0}^{\infty} \int_{0}^{\infty}dr
dr_{1} r^{2} r_{1}^{2}\int_{-1}^{1}d\beta g_{3}({\bf{r,r_{1}}})
[\frac{3}{rr_{1}} \frac{\partial U(r)}{\partial r}\cdot
\frac{\partial U(r_{1})}{\partial r_{1}}\nonumber\\
& &+\frac{r}{r_{1}} \frac{\partial U(r_{1})}{\partial r_{1}}\cdot
\frac{\partial}{\partial r}\left(\frac{\partial U(r)}{r\partial
r}\right) +\frac{r_{1}}{r} \frac{\partial U(r)}{\partial r}\cdot
\frac{\partial}{\partial r_{1}}\left(\frac{\partial
U(r_{1})}{r_{1}\partial r_{1}}\right) \nonumber\\
& &+ rr_{1}\frac{\partial}{\partial r_{1}}\left(\frac{\partial
U(r_{1})}{r_{1}\partial r_{1}}\right)\cdot
\frac{\partial}{\partial r}\left(\frac{\partial U(r)}{r\partial
r}\right)\beta^{2} ],\label{second} \en where
$g_{3}(\bf{r,r_{1}})$ is the triplet correlation function, and
$\beta$ is the cosine of the angle between $r$ and $r_{1}$.

Following the same procedure for  third and second order
relaxation times, $\tau_{3}=\tau_{2}$ (and $i=2$), we come to the
following expression:

\be
D=\frac{k_{B}T}{m}\frac{[\omega^{(4)}-(\omega^{(2)})^{2}]^{3/2}}{[\omega^{(2)}]^{3/2}
[\omega^{(6)}\omega^{(2)}-(\omega^{(4)})^{2}]^{1/2}}.
\label{third} \ee In a more general case with
$\tau_{i}=\tau_{i-1}$, $i=1,2,...$ we can obtain the expression of
diffusion coefficient by the first even frequency moments:
$\omega^{(2)}$, $\omega^{(4)},$... $,\omega^{(2i)}$. The second,
the forth and the sixth frequency moments contained in Eqs.
\Ref{first}, \Ref{second} and \Ref{third} were approximately
obtained for LJ liquid by authors \cite{Tankeshwar}. Using these
values, the diffusion coefficients were calculated with the help
of  Eqs. \Ref{first}, \Ref{second} and \Ref{third}. The comparison
shows that the values of $D$ obtained by Eq. \Ref{second} have the
best agreement with the molecular dynamics data for the whole
studied range of densities and temperatures. The results for
$D^{*}=D(m/\varepsilon\sigma^{2})^{1/2}$ from Eq. \Ref{first} are
presented in Fig. 1 as triangles, from Eq. \Ref{second} are shown
as solid curves and from Eq. \Ref{third} are presented by circles.
We also represented here the results of authors \cite{Tankeshwar}
for comparison with our ones. Namely, the diffusion coefficient
obtained by Ref. \cite{Tankeshwar} from the approximation of the
second-order memory function by the hyperbolic secant are shown as
dotted curves, whereas those obtained in Ref. \cite{Tankeshwar}
from the prescription of Joslin and Gray \cite{Joslin} are shown
by broken and chain curves. From Fig. 1 one can easily see that
the results of Eq. \Ref{second} have a better agreement with the
molecular dynamics data \cite{Heyes,mds} presented by full circles
in more cases than data found of others theories. This result is
of great interest for the study of diffusion phenomena. It
confirms the possibility of equality of relaxation times of  first
and second order memory functions, i. e. $\tau_{1}$ and
$\tau_{2}$. As can be seen from Fig. 1, the approximation with
hyperbolic secant memory show a sufficiently satisfactory
agreement for some cases. Joslin and Gray models yield an
understated results in comparison with molecular dynamics data.
The distinction between these theories and molecular dynamics data
grows with decreasing of density. However, all examined models,
including even Eqs. \Ref{first} and \Ref{third}, begin to
reproduce qualitatively molecular dynamics data as the triple
point of LJ system with the reduced parameters $n^{*}=0.849$,
$T^{*}=0.773$ is approached.

\begin{figure}
\centerline{\epsfig{figure=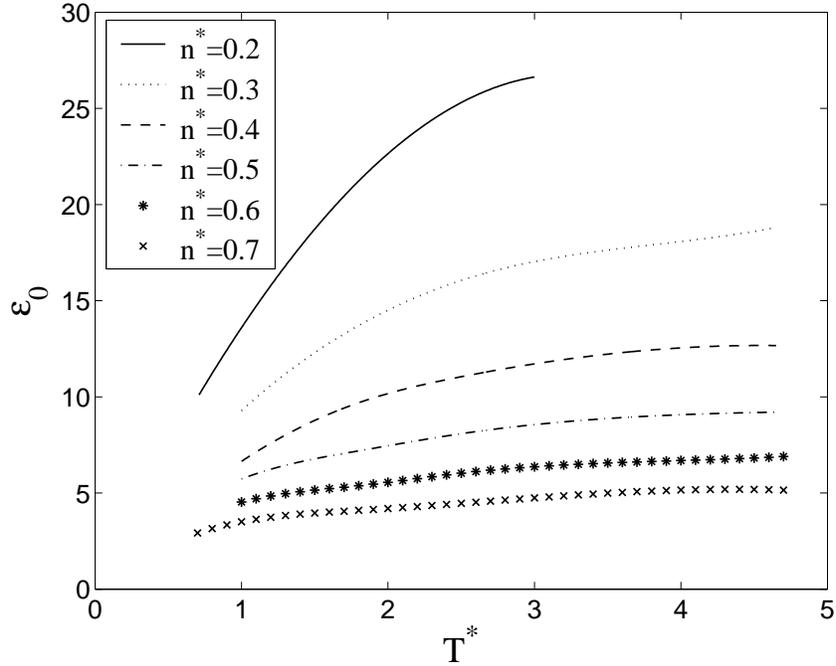,height=9cm,angle=0}}
\caption{Variation of the non-Markovity parameter
$\varepsilon_{0}$ for VACF of LJ liquid with temperature at
different densities. Correspondence between the curves and the
densities is presented in the inset. The behavior of the parameter
$\varepsilon_{0}$ indicate a quantitative change of ratio between
VACF time scale and memory relaxation time. The cross-over from
Markovian ($\varepsilon_{0}\gg 1$) to a quasi-Markovian
($\varepsilon_{0}
>1$) character of the diffusion process at low temperature and high
densities is noteworthy. Referring to the obtained data the
diffusion process is in fact quasi-Markovian in a wide temperature
region at high densities.}
\end{figure}
The next step of our study is estimation of statistical memory
effects in diffusion processes of LJ liquid. It can be
successfully done within the framework of memory function and {\it
dimensionless} non-Markovity parameter \cite{Shurygin1} formalism.
The last one is the ratio of the relaxation time $\tau_{0}$ of
initial correlation function and the relaxation time $\tau_{1}$ of
memory function

\be \varepsilon_{0}=\tau_{0}/\tau_{1}. \label{parameter} \ee

\noindent Using  Eq. \Ref{general} for the relaxation time
$\tau_{0}=\tilde{a}(s=0)$ we obtain

\be \tau_{0}=\frac{mD}{k_{B}T}. \label{tau} \ee

\noindent From the second equality in Eq. \Ref{con_frac} at $s=0$
and Eqs. \Ref{parameter} and \Ref{tau} we come to the final
expression

\be \varepsilon_{0}=\omega^{(2)}\left \lbrack \frac{mD}{k_{B}T}
\right \rbrack^{2}. \label{eps} \ee
\begin{figure}
\centerline{\epsfig{figure=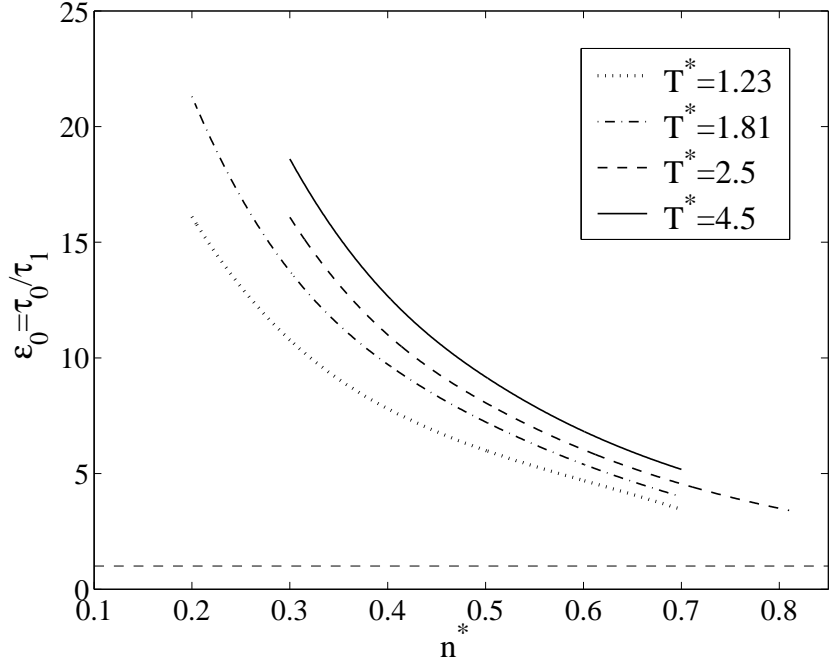,height=9cm,angle=0}}
\caption{Variation of the non-Markovity parameter
$\varepsilon_{0}$ with the reduced density at four different
temperatures $T^{*}=1.23$, $1.81$, $2.5$ and $4.5$ reveal a
nonlinear character of demarkovization at matter densifying. The
horizontal dash line in the bottom of the figure corresponds to
the quantitative value $\varepsilon=1$ (a full non-Markovian
relaxation scenario). The markovization of the process occurs at
decrease of density and increase of temperature; it is accompanied
by increase of the parameter $\varepsilon_{0}$ and partial
disordering of LJ system. Increase of density results in ordering
and demarkovization of diffusion process with sharp reduction of
$\varepsilon_{0}$ values.}
\end{figure}
On the basis of the definition \Ref{parameter} it is possible to
define the spectrum of non-Markovity parameter:
$\varepsilon_{i}=\tau_{i}/\tau_{i+1}$ (see Ref. \cite{Shurygin1}
for a more detail). Then the general form of this parameter for
$i=1,2,3,...$ can be expressed in terms of Mori's coefficients
(static correlation functions) as \be \varepsilon_{i}=\left\{
\begin{array}{rcl}
\displaystyle{\frac{\Omega_{2}^{4}\Omega_{4}^{4}... \Omega_{i+1}^{2}}
{\Omega_{1}^{4}\Omega_{3}^{4}...\Omega_{i}^{4}}}\tau_{0}^{-2}, &\quad{\rm if}& i \quad{\rm is\ odd,}\\
\displaystyle{\frac{\Omega_{1}^{4}\Omega_{3}^{4}...
\Omega_{i+1}^{2}}{\Omega_{2}^{4}\Omega_{4}^{4}...\Omega_{i}^{4}}}\tau_{0}^{2}, &\quad{\rm if}& i\quad{\rm is\ even.}\\
\end{array}
\right.  \label{g_np}
\ee

Obviously, the parameter $\varepsilon_{i}$ obtained by Eq.
\Ref{g_np} estimates the memory effects of the $i$th level
relaxation process, which is described by the memory function
$M_{i}(t)$.

To investigate the memory effects of VACF itself it is necessary
to study the zero point in spectra of the non-Markovity parameter
$\varepsilon_{0}$ calculated by Eq. \Ref{eps}. Therefore, the
parameter $\varepsilon_{0}$ has been calculated for a wide range
of densities and temperatures for LJ liquid. The results of our
calculations are presented in Fig. 2. From this figure one can see
that the non-Markovity parameter always satisfies  condition
$\varepsilon_{0}>1$. This is the evidence of  weak statistical
memory effects in diffusion processes, which is usually observed
in Markovian and quasi-Markovian processes. In this case
$\tau_{1}$ is much smaller than the VACF relaxation time
$\tau_{0}$. The received result helps  to understand the reason of
amazing efficiency of Markovian approximation in the analysis of
diffusion in LJ liquids (see, for example, one of the recent
investigations \cite{Berezhkovskii}). It also enables to estimate
the source of unsatisfactory agreement of Eq. \Ref{first} with
molecular dynamic data. The point is that Eq. \Ref{first} is based
on the assumption of $\tau_{1}=\tau_{0}$ and/or
$\varepsilon_{0}=1$, which is  true for non-Markovian processes
only, while Eqs. \Ref{second} and \Ref{third} are applied both to
Markovian and non-Markovian phenomena. Furthermore, in Fig. 2 we
can see, that the parameter $\varepsilon_{0}$ smoothly increases
with the increase of the reduced temperature
$T^{*}=k_{B}T/\varepsilon$, at the same time it decreases with the
increase of density. So, non-Markovian effects fall off (and/or
Markovity is enhanced)   when the temperature of the system
increases while the  value of the density $n^{*}=n\sigma^{3}$ is
fixed, i. e. markovization of processes is observed, and it
reflects the amplification of randomness during diffusion.
However, a ''saturation '' is attained at a certain value of
temperature, and the values of the parameter $\varepsilon_{0}$
almost cease to change in this case. As may be seen from Fig. 2,
such ''saturation'' is observed at lower temperatures for a more
dense medium. The density $n^{*}$ dependence of the parameter
$\varepsilon_{0}$ is shown in Fig. 3 for the four isotherms. The
behavior of these curves has a nonlinear character. In this figure
we can observe demarkovization, that is, amplification of
regularity and robustness in the system which appears with the
increase of density. For example, for the isotherm $T^{*}=1.81$
the parameter $\varepsilon_{0}$ decreases more than five times
within  density range $0.2 \div 0.7$. Non-Markovian effects are
enhanced and begin to dominate in the vicinity of the triple point
of LJ liquid, where a well-known negative correlation in the
behavior of VACF is observed. The ratio between VACF time-scale
and relaxation time of memory in this phase region is $\sim 2.5
\div 5$. This is a quantitative evidence of considerable memory
effects incipient diffusion process of LJ system near the triple
point, where the density variation is short ranged and of order of
a couple of molecular diameters \cite{cross}.


The results of this work can be summarized as follows. (i) We have
presented a new  approach in calculation of transport
coefficients. The approach is based on the time-scale invariance
idea, developed within framework of Zwanzig-Mori's formalism. Our
theory allows to calculate transport properties in terms of
frequency moments of corresponding correlation functions without
any adjustable parameters. (ii) Three different expressions were
obtained for the diffusion coefficient of LJ liquid and tested
together with other theories and the molecular dynamics data over
a wide range of densities and temperatures. The comparison showed
that the equation obtained from the condition of approximate
equality of relaxation times of the first- and second-order memory
functions has the best agreement with the molecular dynamic data.
This equation includes only the second and the fourth frequency
parameters, which were calculated with a high degree of precision.
(iii) The non-Markovity parameter calculated for the diffusion
phenomena in LJ liquid reveals a Markovian character of thermal
motions of particles. Finally, the values of this parameter
demonstrate the cross-over from Markovian to quasi-Markovian
relaxation scenario at low temperatures and high densities and
allow to estimate quantitatively this cross-over.

The authors acknowledge Professor M. Howard Lee for helpful
discussions and Dr. L.O. Svirina for technical assistance. This
work was partially supported by the the Russian Foundation for
Basic Research (Grants No. 02-02-16146), the Russian Ministry of
Education (Grants No. 03-06-00218a, A03-2.9-336) and RHSF (Grant
No. 03-06-00218a).


\begin{thebibliography}{99}
\bibitem{expon} It is important to note that the initial time correlation function may be
presented as linear combination of exponential functions
$a(t)=\sum_{i}b_{i}e^{-t/\tau_{i}}$ at generalization of closure
to time-dependence at this relaxation level, $M_{2}(t) \approx
M_{1}(t)$ (detailed discussion of this finding will be published
soon).
\bibitem{mds} As an ''experimental'' data we use results of Heyes \cite{Heyes},
which are well known and applied often for different theories
testing \cite{Tankeshwar,Rah}. The VACF in these molecular
dynamics simulations is obtained by averaging over all atoms in
the system and over several time origins; the integral over all
time of the VACF yields the diffusion coefficient \cite{Hansen}.
For a more detailed description of the molecular dynamics
simulations, the reader is referred to Ref. \cite{Heyes}.
\bibitem{cross} From the physical point of view the cross-over
from a Markovian to a quasi-Markovian character (of the diffusive
process) at low temperatures and high densities is not a suprise.
For a more quantitave description of this cross-over it should be
recognized that the range of negative correlation is observed in
the vicinity of the triple point of LJ liquid. As known, this
correlation region is analogical (but a more considerable) to a
such range for the hard spheres system.
\end{thebibliography}
\end{document}